\documentclass[letterpaper,12pt]{article}
\pdfoutput=1

\usepackage{jheppub}

\usepackage{afterpage}

\usepackage[lofdepth,lotdepth,caption=false]{subfig}
\usepackage{fancyhdr}
\usepackage{hyperref}
\usepackage{amsmath, amssymb, graphicx,amsfonts}
\usepackage{mathrsfs}
\usepackage{xspace}
\usepackage{braket}
\usepackage{color}
\usepackage{setspace}
\usepackage{mathtools}
\usepackage[dvipsnames]{xcolor}
\usepackage{float}
\usepackage{afterpage}

\usepackage{dsfont}
\usepackage{tikz}
\usepackage{ulem}
\usepackage{cancel}

\usepackage{alltt}
\allowdisplaybreaks

\def\({\left(}
\def\){\right)}
\def\[{\left[}
\def\]{\right]}

\newcommand{\be}{\begin{equation}}
\newcommand{\ee}{\end{equation}}

\newcommand{\ban}[1]{\begin{align}#1\end{align}}

\newcommand{\corr}[1]{\left< #1\right>}

\newcommand{\pp}[2]{\begin{pmatrix}#1 \\#2 \end{pmatrix}}

\title{Numerical Methods for Handlebody Phases}

\author{Jason Wien}

\affiliation{Department of Physics, University of California, Santa Barbara, CA 93106, USA}

\emailAdd{jswien@gmail.com}

\abstract{
We review methods used in recent works for constructing handlebody solutions of Einstein's equations in 2+1 dimensions. Additionally, we provide a \textit{Mathematica} package for computing the action and the boundary moduli of these solutions in a canonical conformal frame.
}

\makeatletter
\def\@fpheader{\relax}
\makeatother

\begin{document}
\maketitle

\section{Introduction}
\label{section:intro}

The AdS/CFT correspondence has proven to be a fruitful avenue for probing various aspects of quantum gravity \cite{maldholo,wittenholo}. In particular, classical black hole spacetimes of non-trivial topology in three dimensions have been used to study parition functions and operators in 2d holographic CFTs \cite{MRW,toruspaper} as well as probe the entanglement structure of states in such theories \cite{MBW1, MBW2,cones}. These solutions arise as saddle points of the Euclidean Einstein-Hilbert action with boundary conditions given $\partial M = X$, where $X$ is a compact Riemann surface. One can think of constructing them by filling in various cycles of $X$, and so they are often referred to as handlebody phases \cite{Brill1, Brill2, Skenderis, Krasnov1, Krasnov2}. 

We review some of the tools used in these works \cite{MRW,cones,toruspaper} to study these handlebody solutions of Einstein's equations. These techniques include the finite element method (FEM) for numerically solving differential equations as well as the mathematical framework of Schottky uniformization for characterizing Riemann surfaces. The focus will be on useful formulas for implementing these calculations, and as such we have attached a \textit{Mathematica} package to the electronic version of this work that implements many of these tools.

The outline of this review is as follows. First, we review the aspects of finite element methods used in these constructions. Next, we give a rough overview of Schottky uniformization, focusing on explicitly writing down a uniformization for a given Riemann surface that describes a desired handlebody phase. We additionally give explicit formulas for computing the regularized action of these phases reduced by certain symmetries, which we conveniently encode in the attached \textit{Mathematica} package. Finally, we give a simple example to illustrate the concepts and techniques described.

\section{Finite Element Methods}

Finite element methods are numerical methods for solving differential equations which involve discretizing the domain with a set of finite ``elements'' \cite{FEMgentle,FEMlecture}. We will restrict our attention exclusively to equations in two dimensions of the form
\ban{
\nabla^2 u(x,y) + f(x,y) u(x,y) = g(x,y) \, \label{FEMeq} .
}
In order to solve this equation, we will discretize our solution space and convert this equation into a finite dimensional matrix equation, which we can then easily solve by algebraic methods. 

\subsection{Discretization of the domain}
\begin{figure}[h!]
\centering
\includegraphics[width=0.45\textwidth]{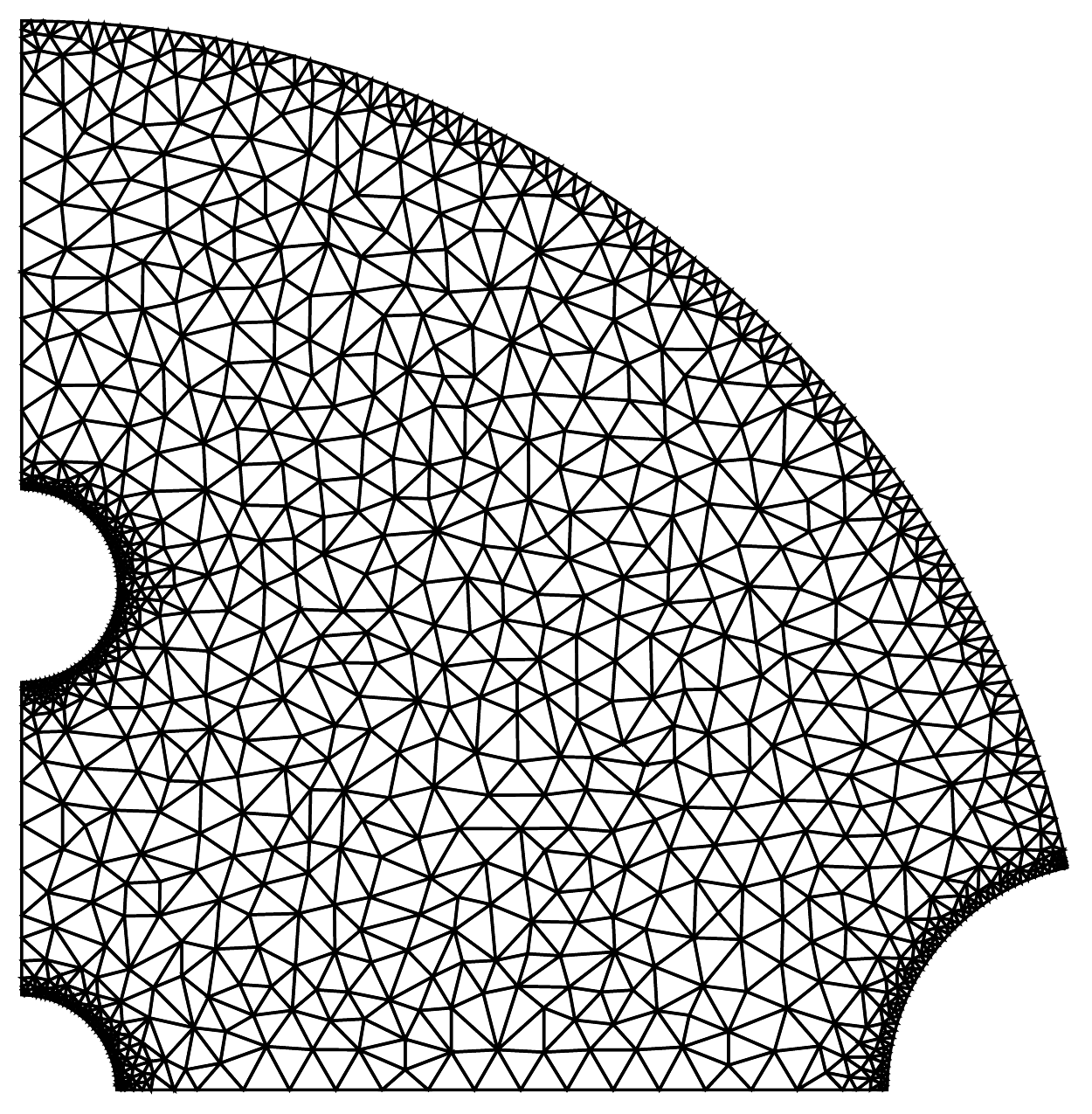}
\caption{An example FEM mesh used in \cite{toruspaper}. \label{mesh}
}
\end{figure}

First, we discretize the domain $D$ with a mesh made up of triangular elements. We will use elements with six nodes: one on each vertex and one on the midpoint of each edge. An example of a valid triangulation for a domain used in \cite{toruspaper} is shown in figure \ref{mesh}. 
We generate meshes using \textit{Mathematica}'s built-in \verb!ToElementMesh! function. Note that for numerical convenience, we approximate curved boundaries of $D$ by a large number of straight segments. We can estimate the error introduced by computing the length of $\partial D$ using the mesh and comparing it to the true value. The error introduced by this approximation can easily be made smaller by including more nodes on the boundary, and we always choose a sufficient number of nodes so that this error is always sub-leading. 

Given a valid mesh, we can define our solution space as the Sobolev space of piecewise continuous second-order polyomials spanned by the set of functions $\psi_i$ on $D$ such that $\psi_i(n_j) = \delta_{ij} $. That is, we parameterize our solution space with a basis of second order polynomials such that $\psi_i$ is one on node $n_i$ and vanishes on all other nodes. In this way we can approximate any function as
\ban{
u \approx \sum_{i=1}^N u_i \psi_i \, , \label{FEMapprox}
}
where $u_i = u(n_i)$ and $N$ is the number of nodes in the mesh. We can improve this approximation by increasing the number of elements in the mesh. 

\begin{figure}[h!]
\centering
\includegraphics[width=0.65\textwidth]{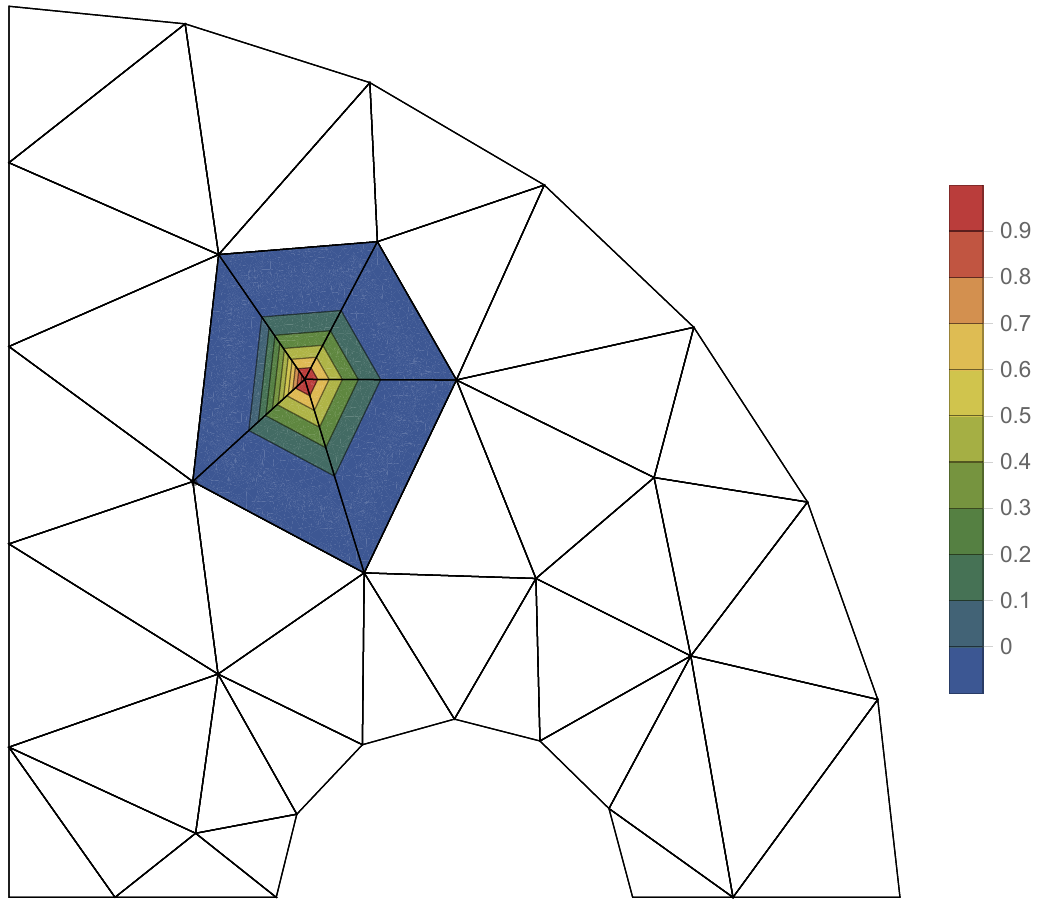}
\caption{A contour plot of $\psi_i$ for a particular $n_i$ and FEM mesh. Note that $\psi_i$ is $1$ on $n_i$, $0$ on $n_j \neq n_i$, and non-vanishing only in the highlighted neighborhood $\mathscr N_i$. \label{neighborhood}
}
\end{figure}

Note that $\psi_i$ is non-vanishing only on the set of elements containing $n_i$. In the discussion below and in the attached code, we refer to such a set as the ``neighborhood'' of $n_i$, and we can simplify some of the computations by restricting only to the appropriate neighborhood. We plot an example $\psi_i$ and highlight its associated neighborhood in figure \ref{neighborhood}. 

\subsection{Solving the differential equation}

To convert the equation \eqref{FEMeq} to a matrix equation, we can integrate both sides against an arbitrary $\psi_i$. 
\ban{
\int_D \nabla^2 u \, \psi_i  + \int_D f \, u\, \psi_i = \int _D  g \, \psi_i \, .
}
Integrating by parts gives the equation
\ban{
\int _{\partial D} \nabla_n u\,  \psi_i  - \int_D \nabla u \cdot \nabla \psi_i  + \int _D f\, u \, \psi_i = \int_D g\,  \psi_i\, .
}
Finally we can use the approximation eq. \eqref{FEMapprox} to convert this equation into a matrix equation:
\ban{
\sum_{j=1}^N u_j \left[ \int_{\partial D} \psi_i \nabla_n \psi_j \right]  - \sum_{j=1}^N u_j  \left[ \int_{ D} \nabla \psi_i \cdot \nabla \psi_j \right] + \sum_{j=1}^N u_j  f_j  \left[ \int_{ D} \psi_i \psi_j \right]  =  \sum_{j=1}^N g_j  \left[ \int_{ D} \psi_i \psi_j \right] \, . 
}
This equation is a bit ugly, but we can clean it up by introducing the following notation:
\ban{
M_{ij} =\int_{ D}  \psi_i  \psi_j \,  \hspace{1cm}W_{ij} = \int_{ D} \nabla \psi_i \cdot \nabla \psi_j  \hspace{1cm} K_{ij} = \int_{\partial D} \psi_i \nabla_n \psi_j
}
where $M$ and $W$ are often called the ``mass'' and ``stiffness'' matrices respectively. Note that $K_{ij}$ is non-zero only when both $n_i$ and $n_j$ are on the boundary.\footnote{Additionally it is often possible to rewrite $K_{ij}$ in a simpler manner using the boundary conditions. We do so in the applications of FEM in \cite{cones,toruspaper} and later in this section.} With these definitions we can write our equation as 
\ban{
\sum_{j=1}^N K_{ij} u_j   - \sum_{j=1}^N W_{ij} u_j  + \sum_{j,k=1}^N M_{ij} (f_{j}\delta_{kj}) u_k   =  \sum_{j=1}^N M_{ij} g_j  \notag \\
\sum_{j=1}^N \left[ K_{ij} - W_{ij}  + \sum_{k=1}^N M_{ik} (f_{k}\delta_{kj})\right] u_j  =  \sum_{j=1}^N M_{ij} g_j  \, , \label{matrixEQ}
}
which now takes the form of a matrix equation $A \cdot \vec u = \vec b$. We can easily solve this equation using the \verb!LinearSolve! function in \textit{Mathematica}, after appropriately enforcing the boundary conditions. 

We will exclusively consider boundary conditions which can be converted into a Neumann-type form, as in \cite{MRW,cones,toruspaper}. That is, we only consider cases where we can rewrite $K_{ij}$ using the boundary conditions $\nabla_n u = f$ in the manner
\ban{
\sum_{j=1}^N K_{ij}u_j =\int_{\partial D} \psi_i \nabla_n u =\int_{\partial D} \psi_i f =\sum_{j=1}^N \int_{\partial D} \psi_i \psi_j f_j \, . 
}
In this way, we have converted the term $K_{ij}$ in our matrix equation into a source term given by $\sum_{j=1}^N C_{ij} f_j$ where 
\ban{
C_{ij} = \int_{\partial D} \psi_i \psi_j \, .
}
This new source term enforces the appropriate boundary conditions, and so no further modifications need to be made to eq. \eqref{matrixEQ} to ensure the solution obeys them. The modified equation is given by 
\ban{
\sum_{j=1}^N \left[- W_{ij}  + \sum_{k=1}^N M_{ik} (f_{k}\delta_{kj})\right] u_j  =  \sum_{j=1}^N\left[ M_{ij} g_j -C_{ij} f_j \right]\, .\label{matrixEQ2}
}

\subsection{Computation of matrix elements}

\begin{figure}[h!]
\centering
\includegraphics[width=0.45\textwidth]{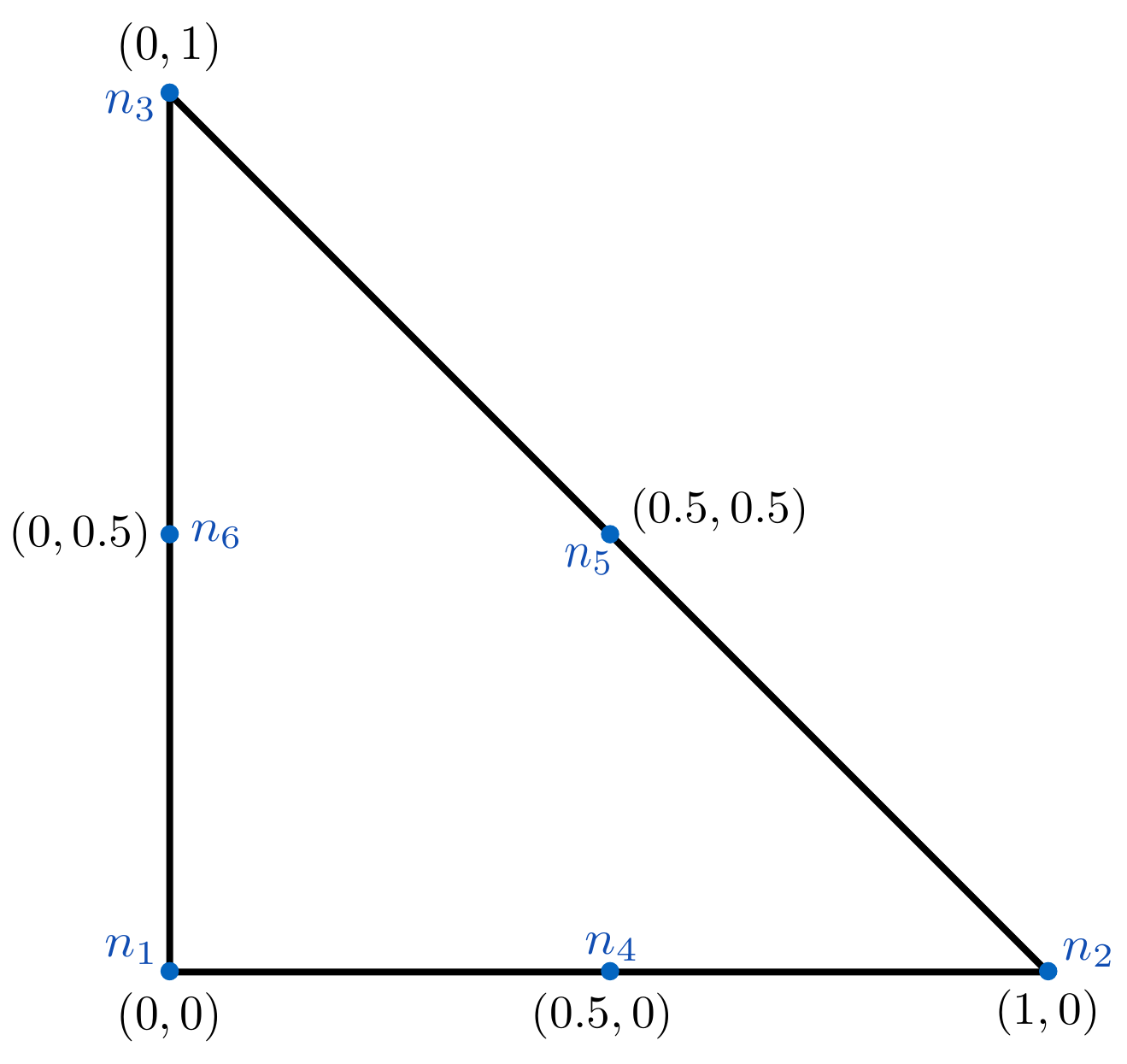}
\caption{The unit reference element with nodes labeled. \label{elem}
}
\end{figure}

In practice, we can compute the matrices $M$, $W$, and $C$ by deriving an analytic formula based on a unit ``reference element'' $R$ with vertices at $(0,0)$, $(0,1)$, and $(1,0)$ as drawn in figure \ref{elem}. 
Note that as $\psi_i$ is non-vanishing only in the neighborhood of $n_i$ (denoted by $\mathscr N_i$) we can write
\ban{
M_{ij} = \int_{ D}  \psi_i  \psi_j = \int_{ \mathscr N_i \cap \mathscr N_j} \psi_i \psi_j = \sum_{E \in \mathscr N_i \cap \mathscr N_j} \int_E \psi_i \psi_j\, .
}
Therefore we can decompose $M_{ij}$ (and similarly $W_{ij}$ and $C_{ij}$) as a sum of integrals over elements in $\mathscr N_i \cap \mathscr N_j$. It will be useful then to derive analytic formulas for the following integrals over an arbitrary element $E$ specified by the coordinates of its vertices:
\ban{
m_{ij} =\int_{E}  \psi_i  \psi_j \,  \hspace{1cm}w_{ij} = \int_{E} \nabla \psi_i \cdot \nabla \psi_j  \hspace{1cm}  {c_{ij}(s)} = \int_{s} \psi_i  \psi_j  \, ,
}
where $E$ is assumed to contain nodes $n_i$ and $n_j$ and $s$ is a particular boundary segment of $E$. First we can compute the value of these integrals on the unit reference element, then transform to an arbitrary element $E$ using an appropriate change of coordinates.\footnote{One can also compute these integrals using Gaussian quadrature rules as in \cite{MRW}, but we choose to eliminate the need to compute any analytic derivatives of the $\psi_i$.} Quantities associated with the reference element we denote by a superscript $R$. 

First we can write $\psi_i$ for the reference element:
\ban{
\begin{array}{lll}
\psi_1^{(R)} = (x+y-1)(2x+2y-1)\, , & \psi_2^{(R)} = x (2x-1)\, , & \psi_3^{(R)} = y (2y-1)\, ,\\
\psi_4^{(R)} = 4x (1-x-y) \, ,&\psi_5^{(R)} =4xy \, ,& \psi_6^{(R)} = 4y (1-x-y)\, .
\end{array}
}
One can easily see that these functions are second order polynomials that satisfy $\psi_i(n_j) = \delta_{ij} $ as required. Additionally, given these expressions we can analytically compute the 36 matrix elements of $m^{(R)}_{ij}$, $w^{(R)}_{ij}$, and each of the three $c_{ij}^{(R)}(s)$. 

To transform from the reference element to an arbitrary element with nodes $n_i = (x_i, y_i)$, we can perform the coordinate transformation
\ban{
\pp {x'}{y'} = \begin{pmatrix} x_2-x_1 & x_3 - x_1 \\ y_2 -y_1& y_3 - y_1 \end{pmatrix}\pp x y + \pp{x_1}{y_1} \, .
}
Using the standard change of basis formulas for integrals and derivatives, we can derive analytic expressions for the matrix elements as functions of the vertices $(x_i,y_i)$ of $E$:
\ban{
m_{ij}&= |J|  \footnotesize
\begin{bmatrix}
 \frac{1}{60} & -\frac{1}{360} & -\frac{1}{360} & 0 & -\frac{1}{90} & 0 \\
 -\frac{1}{360} & \frac{1}{60} & -\frac{1}{360} & 0 & 0 & -\frac{1}{90} \\
 -\frac{1}{360} & -\frac{1}{360} & \frac{1}{60} & -\frac{1}{90} & 0 & 0 \\
 0 & 0 & -\frac{1}{90} & \frac{4}{45} & \frac{2}{45} & \frac{2}{45} \\
 -\frac{1}{90} & 0 & 0 & \frac{2}{45} & \frac{4}{45} & \frac{2}{45} \\
 0 & -\frac{1}{90} & 0 & \frac{2}{45} & \frac{2}{45} & \frac{4}{45} \\
\end{bmatrix}
 \notag \\
w_{ij} &= \frac 1 {6|J|}  \footnotesize
\begin{bmatrix}
 3 \, \chi _{23} & \xi _3 & \xi _2 & -4 \, \xi _3 & 0 & -4 \, \xi _2 \\
 \xi _3 & 3 \, \chi _{13} & \xi _1 & -4 \, \xi _3 & -4 \, \xi _1 & 0 \\
 \xi _2 & \xi _1 & 3 \, \chi _{12} & 0 & -4 \, \xi _1 & -4 \, \xi _2 \\
 -4 \, \xi _3 & -4 \, \xi _3 & 0 & 4 \left(\chi _{12}+\chi _{13}+\chi _{23}\right) & -8 \, \xi _2 & -8 \, \xi _1 \\
 0 & -4 \, \xi _1 & -4 \, \xi _1 & -8 \, \xi _2 & 4 \left(\chi _{12}+\chi _{13}+\chi _{23}\right) & -8 \, \xi _3 \\
 -4 \, \xi _2 & 0 & -4 \, \xi _2 & -8 \, \xi _1 & -8 \, \xi _3 & 4 \left(\chi _{12}+\chi _{13}+\chi _{23}\right) \\
\end{bmatrix}\, ,
}
where we have used the notations 
\ban{
|J| &=  | \left(x_3-x_2\right) y_1+\left(x_1-x_3\right) y_2+\left(x_2-x_1\right) y_3 |\notag \\
\xi_1 &= (x_1-x_2)(x_1-x_3) +(y_1-y_2)(y_1-y_3) \notag\\ \xi_2 &= (x_2-x_1)(x_2-x_3) +(y_2-y_1)(y_2-y_3) \notag\\ \xi_3 &= (x_3-x_1)(x_3-x_2) +(y_3-y_1)(y_3-y_2)\notag\\
\chi_{12} &= (x_1-x_2)^2+(y_1-y_2)^2\notag\\ \chi_{23} &= (x_2-x_3)^2+(y_2-y_3)^2\notag\\ \chi_{13} &= (x_1-x_3)^2+(y_1-y_3)^2 \, .
} 
For $c^{(s)}_{ij}$ we can write the matrix elements as
\ban{
&c^{(s)}_{a_s a_s} = c^{(s)}_{b_s b_s} = 2/15 |s| \notag\\
&c^{(s)}_{a_s d_s} = c^{(s)}_{b_s d_s} = 1/15 |s|\notag\\
&c^{(s)}_{a_s b_s} =-1/30 |s|\notag\\
& c^{(s)}_{d_s d_s} = 8/15|s| \, ,
}
where all matrix elements not implied by $i\leftrightarrow j$ symmetry vanish and segment $s$ extends between nodes $a_s$ and $b_s$ with midpoint $d_s$, and $|s|$ is the Euclidean length of segment $s$. Using these formulas provides an efficient way to compute $M_{ij}$, $W_{ij}$, and $C_{ij}$ and, then numerically solve a given differential equation in terms of the matrix equation eq. \eqref{matrixEQ2}. 

%
%

\section{Handlebody Phases}

All solutions of vacuum Einstein's equations with negative cosmological constant in 2+1 dimensions are quotients of AdS$_3$. These solutions provide a rich set of spacetimes for probing holography, as we are able to construct geometries with non-trivial topology simply by taking quotients. These geometries often arise as the gravitational duals of CFT states in two dimensions defined via a Euclidean path integrals. For a state defined as a path integral over a genus $g$ Riemann surface $X$, the associated gravitational path integral with boundary conditions $\partial M = X$ has a set of Euclidean saddles which we can characterize by specifying a set of $g$ cycles on the boundary to be made contractible in the bulk. We refer to these saddles as handlebody phases, which have been extensively studied in \cite{Brill1, Brill2, Skenderis, Krasnov1, Krasnov2, MRW}, and which are the focus of this section.

In this section, we review methods for constructing these handlebody phases and evaluating their actions. We focus on practical tools and formulas for doing computations, and we refer the reader to \cite{toruspaper} and the various references for more details on the rich mathematical theory underlying these methods. In particular, we will show how to compute the regularized Einstein Hilbert action in the conformal frame where $R_\text{bndy} = -2/\ell^2$, and we will set $\ell=1$.

\subsection{Schottky Uniformization}

We can construct a convenient representation of a handlebody phase, called a Schottky uniformization, by starting with the boundary Riemann surface $X$ of genus $g$. To specify a handlebody phase, we need to choose a set of $g$ independent and non-intersecting cycles to be made contractible in the bulk. For example, given a basis $\{\alpha_i, \beta_j\}$ of the homotopy group of $X$ such that $\alpha_i  \beta_j = \delta_{ij}$ and $\prod_i \alpha_i^{-1}\beta_i^{-1}\alpha_i  \beta_i  = 1$, we can choose the set of $g$ cycles $\{\alpha_i\}$, the cycles $\{\beta_i\}$, or any set of cycles given by the image of $\{\alpha_i\}$ under an element of the mapping class group. 

Having chosen a set of $g$ cycles, we now cut open the Riemann surface along each cycle and label each side of the cut $C_i$ and $C_i'$. The resulting surface is a Riemann sphere punctured by $2g$ circles that come in pairs. We can project this sphere into the complex plane, resulting in a Schottky domain for $X$. It is often useful to make sure that certain reflection and rotational symmetries of $X$ are preserved along the way, although it is sometimes not possible to preserve all such symmetries. Alternatively, one can begin with $2g$ circles in the complex plane and then reverse engineer the corresponding surface $X$ and handlebody phase, although we found this process to be more difficult in practice.\footnote{There is an additional complication that sometimes the symmetries of $X$ act in a non-trivial way on the Schottky uniformization, so determining the bulk geometry on a particular symmetry slice of the boundary can be difficult.} 

The region in $\mathbb C$ exterior to all the $C_i$ and $C_i'$ can be taken as a fundamental domain $D$ for the surface $X$. As $C_i$ and $C_i'$ are the same cycle on $X$, we can recover $X$ from the Schottky uniformization by taking the quotient by the subgroup of M\"obius transformations $\corr{L_i}$, where each $L_i$ maps the interior of $C_i$ to the exterior of $C_i'$.

The Schottky domain resulting from this construction describes a bulk phase in which the initial cycles chosen on the boundary are contractible in the bulk. If we consider the half-plane model of $\mathbb H^3$ with the complex plane as its boundary,\footnote{We remind the reader that Euclidean AdS$_3$ is $\mathbb H^3$ or three dimensional hyperbolic space.} we can extend the identifications on the boundary into the bulk along geodesic hemispheres. That is, the quotient group acts in the bulk by identifying the hemispheres anchored on $C_i$ and $C_i'$. In this way, the cycles homologous to $C_i$ on the boundary are contractible in the bulk, as they may be lifted off the boundary along the corresponding hemisphere and shrunk down to a point. The dual cycles running between $C_i$ and $C_i'$ remain non-contractible. Therefore, we have successfully described the handlebody phase with the requisite boundary cycles contractible in the bulk.

One way to characterize a handlebody phase is by the topology of a particular slice through the bulk, often corresponding to a moment of time-reflection symmetry. When this slice is fixed by a reflection symmetry of the boundary $X$, we can compute the topology using the following formula: 
\ban{
g_\text{slice} = \frac 12 (n-b+1)\, , \label{sliceG}
}
where $b$ is the number of disconnected boundaries of the slice and $n$ is the number of pairs of circles that lie on the slice. Note that the assumption of reflection symmetry ensures that either both circles of a pair lie on the slice or neither do. For example, a slice intersecting $2$ pairs of circles that divides the boundary into $3$ disconnected circles has no topology in the interior, and so this slice describes a simple three boundary wormhole.

To compare the gravitational action between different phases, we must numerically solve for a standard conformal frame on the boundary and regularize the action. Additionally, we must be sure to compare phases with the same boundary $X$, and so we will need to compute the moduli of the boundary for each phase, and match the moduli between phases. This process is computationally intensive, but we may sometimes use a heuristic to get a rough understanding of the phase diagram. 

In general the phase with minimal action will be the one in which the total length of the boundary cycles made contractible is minimized.\footnote{Note that we have fixed the conformal frame of the boundary to be $R_\text{bndy}=-2$.} Note that for many phases there is not a unique choice of $g$ cycles that yield that phase, and so when applying the heuristic one must choose the choice of $g$ cycles that yields the minimal action. We can summarize this heuristic simply as: \textit{``Shorter cycles are more likely to pinch off than longer cycles''}.
While this heuristic does not hold exactly (in fact we can construct cases where it fails), it is true approximately in the sense that as boundary cycles get longer the phase in which they are contractible becomes more subdominant. In this way, this heuristic is a useful shortcut for determining the general structure of the phase diagram.  

\subsection{The boundary metric}
 
In order to fully specify the boundary Riemann surface $X$ and the corresponding handlebody phase, along with the set of contractible cycles we need to additionally specify the $3g-3$ moduli of the boundary. In the cases we consider, some of the moduli are fixed by symmetry, while others are computed by evaluating the lengths of certain geodesics on the boundary. Therefore, we need to specify a boundary metric before we can fully match a Schottky uniformization with its Riemann surface $X$. As detailed in \cite{Krasnov1}, to properly renormalize the gravitational action, we should choose a conformal frame on the boundary in which $R_\text{bndy}=-2$. As all metrics in 2d are conformally flat, we can write in general
\ban{
ds^2 = e^{\phi(w)} |dw|^2\, ,
}
where $\phi(w)$ is an arbitrary function for which we will solve. The regularity of the metric under the quotient by the $L_i$ imposes the following boundary conditions on $\partial D$:
\ban{
\phi(L_i(w)) = \phi(w) - \frac 12 \log |L'_i(w)|^2\, . \label{bcs}
}
Additionally, the requirement $R_\text{bndy}=-2$ yields the Liouville equation for $\phi$:
\ban{
\nabla^2 \phi = e^{2 \phi}\, . \label{Leqn}
}

In all cases we consider, the circles $C_i$ and $C_i'$ are fixed point sets of a symmetry of $D$ given by inversion through some circle in the complex plane. Using polar coordinates $(r_I,\theta_I)$ centered on the circle of inversion with radius $R_I$, invariance of the metric under this symmetry requires that
\ban{
\phi(R^2_I /r_I , \theta_I ) = \phi(r_I, \theta_I)+ \log(r_I^2/R_I^2)\, .
}
Differentiating with respect to the unit normal $\hat r_I$ we find
\ban{
\partial_{r_I}\phi(R_I^2/r_I, \theta)= - \frac{r_I^2}{R_I^2}\left( \partial_{r_I} \phi(r_I, \theta_I) + \frac 2 {r_I}\right)\, . \label{inversionSym}
}
Evaluating this equation on $r_I = R_I$ we have the simple formula that on $C_I$ 
\ban{
\left . \partial_{R_I} \phi \right|_{C_I} = - \frac 1 {R_I} \, . \label{INVbc}
}

In fact, we can show that when $C_i$ and $C_{i'}$ are related by an involution symmetry, this equation also holds on $C_i$ and $C_{i'}$. First, we consider $C$ and $C'$ as concentric circles centered at the origin with radii $\lambda$ and $1/\lambda$ respectively with $\lambda >1$, and $L(w) = w/\lambda^2$. The domain $D$ is the region between the circles\footnote{Note that the ``outside'' of $C$ is the region including the origin.}. From the boundary conditions eq. \eqref{bcs} we have 
\ban{
\frac 1 {\lambda^2} \partial_r \phi(1/\lambda) = \partial_r \phi(\lambda)\, .
}
Additionally, $C$ and $C'$ are related by inversion through the unit circle, and so by eq. \eqref{inversionSym} we have
\ban{
\partial_r \phi(1/\lambda) = - \lambda^2 \left(\partial_r \phi(\lambda) + \frac 2 \lambda \right)\, .
}
Solving these two equations and noting $\nabla_n = \pm \partial_r$ for $C$ and $C'$ respectively we have
\ban{
\left. \nabla_n \phi\right|_C = - \frac 1 \lambda \hspace{1cm} \left. \nabla_n \phi\right|_{C'} = \lambda \, . \label{Cbc}
}
Or in general we have 
\ban{
\left. \nabla_n \phi \right|_{C_i} = \frac {\sigma}{R_i} \label{FEMbc}\, ,
}
where $\sigma_i = \pm 1$ for $D$ outside or inside of $C_i$ respectively and $R_i$ is the radius of $C_i$. 

To show that the condition eq. \eqref{FEMbc} holds whenever $C_i$ and $C_{i'}$ are related by an involution symmetry, we can perform a M\"obius transformation to move the unit circle to the appropriate circle of involution. Let the appropriate transformation be given by 
\ban{
w' = \frac{a\,w + b}{c\,w+d} \label{Mtrans}\, .
}
Under this transformation we have
\ban{
\vec\nabla' \left(\phi(w') -\frac 12 \log \left|\frac{bc-ad}{(a-c\, w')^2}\right|^2\right)= J^{-1}\cdot \vec \nabla \phi(w)\, ,
}
where $J$ is the Jacobian of the transformation eq. \eqref{Mtrans}. As on $C$ and $C'$ we know $\vec \nabla \phi(w)$ from eq. \eqref{Cbc} and we can compute $J$, we can solve this equation for $\vec \nabla {}'\phi(w')$. Taking the inner product with the normal vector on the image of $C$ and $C'$ under the coordinate transformation yields eq. \eqref{FEMbc}.  

We can now solve eq. \eqref{Leqn} using the Newton-Raphson algorithm and the finite element methods described in the previous section. First, we write $\phi= \phi_{(n)} + \delta \phi_{(n)}$ and expand the Liouville to first order in $\delta\phi_{(n)}$:
\ban{
\nabla^2 \delta\phi_{(n)} - 2 e^{2\phi_{(n)}}\delta\phi_{(n)} = -\left(\nabla^2 \phi_{(n)} - e^{2\phi_{(n)}} \right)\, .
}
With the assumption that all $C_i$ and $C_{i'}$ are related by a $\mathbb Z_2$ symmetry of the domain, we can rewrite the boundary conditions eq. \eqref{bcs} as Neumann-type conditions eq. \eqref{FEMbc}. In the manner discussed in the previous section, we can enforce these boundary conditions by introducing a source term in the integral form of our differential equation:\footnote{Note that in the last term the orientation $\sigma_i$ is absorbed into the orientation of $d\theta_i$ in the manner described in the next section.}
\ban{
- \int_D \nabla \psi\cdot \nabla \delta \phi_{(n)}- 2 \int_D \psi \, e^{2\phi_{(n)}} \, \delta\phi_{(n)} = \int_D \nabla \psi \cdot \nabla \phi_{(n)} + \int_D \psi \, e^{2\phi_{(n)}} + \sum_{i}\frac {1}{R_i} \int_{\partial D_i} {\psi}\, d \theta_i \, . \label{inteq}
}
This equation is now in the form to apply the formulas from the previous section.

Further, we can often use the symmetries of the Schottky uniformization to reduce $D$ down to a reduced domain $\tilde D$. In all cases we consider, we use at least one reflection symmetry to reduce $D$, and without loss of generality we can choose for this reflection symmetry to act as inversion through the unit circle. Therefore, we choose to always work with a finite domain $\tilde D$. Note that the boundary conditions on the unit circle are fixed by eq. \eqref{INVbc}, and are accounted for by the final term in eq. \eqref{inteq}. 

Using FEM to discretize this equation, we then can solve the appropriate matrix equation for $\delta\phi_{(n)}$. Then, we update our solution to $\phi_{(n+1)} = \phi_{(n)}+ \delta \phi_{(n)}$ and solve a similar equation for $\delta\phi_{(n)}$. Starting with an initial seed of $\phi_{(0)}=0$, we repeat this process until $||\delta\phi_{(n+1)}||_\infty<10^{-10}$ or another desired accuracy. 

Given the solution for $\phi$, we can use the metric to numerically compute the lengths of all segments of $\partial \tilde D$. However, to compute the lengths of geodesics that do not make up $\partial \tilde D$ we must use a different method. We note that the region $\tilde D$ with $R_\text{bndy}=-2$ can be represented as a region in $\mathbb H^2$, and so if we can construct this region we can use the known analytic properties of $\mathbb H^2$ to compute the lengths of geodesics. Given a region $\tilde D$ with boundary segments $\partial \tilde D_i$ given by geodesics that meet at right angles,\footnote{This condition will be guaranteed by our symmetry requirements.} we can construct a corresponding region in $\mathbb H^2$ by the following algorithm. First, we start with an arbitrary geodesic segment of length $|\partial \tilde D_1|$. Next, we solve for the geodesic in $\mathbb H^2$ that intersects it orthogonally, and we follow that geodesic for length $|\partial \tilde D_2|$. We continue this process until we have represented all boundary segments and form a closed region. Using this region in $\mathbb H^2$, we can now solve for the lengths of geodesics using well known formulas. In this way, we can compute all the remaining moduli of the boundary $X$. 

\subsection{The bulk action}

We can now compute the Einstein-Hilbert action for the associated handlebody phase. In terms of the field $\phi$ it was shown in \cite{MRW} that the regularized action is given by 
\ban{
I = - \frac c{24\pi} \left[ I_\text{TZ}[\phi] - A - 4 \pi (g-1)(1-\log 4 \rho_0^2)\right]\, , \label{fullaction}
}
where $A$ is the area of the boundary, $c=3/2G_N$ is the central charge of the dual CFT, and $\rho_0$ is the radius of the sphere for which the partition function is one, and we set $\rho_0=1$.  Additionally defining $R_i$ to be the radius of $C_i$ and $\Delta_i$ as the distance between the center of $C_i$ and the point $w_\infty^{(i)}$ mapped to $\infty$ by $L_i$, we have,
\ban{
I_{TZ}[\phi] = \int_D d^2 w\left( \left(\nabla \phi\right)^2 + e^{2\phi} \right) + \sum_i \left(\int_{C_i} 4 \phi\, d\theta_\infty^{(i)}  - 4 \pi \log \left |R_i^2 - \Delta_i^2 \right|\right)\,, \label{TZaction}
}
where $\theta_\infty^{(i)}$ is the angle measured from the point $w_{\infty}^{(i)}$. In the rest of this section, we use our assumption of symmetries to simplify this action and derive useful formulas. 

First, we note that on shell we have the relation
\ban{
A =  \int_D d^2 w \, e^{2\phi}\, 
}
and therefore the term $A$ in eq. \eqref{fullaction} cancels part of the integration in eq. \eqref{TZaction}. Further, we can reduce the remaining ntegral over $D$ to integrations over $\tilde D$ using the various inversion and reflection symmetries. Using the relation eq. \eqref{inversionSym} we have
\ban{
\int_D d^2 w \left(\nabla \phi\right)^2  = 2 \int_{\tilde D}d^2 w  \left[\left(\nabla \phi\right)^2 + \frac{2}{r_I}\partial_{r_I}\phi + \frac 2{r_I^2} \right]\, .
}
In practice, we only use reflections and inversion through the unit circle to reduce the Schottky domain.\footnote{Note that for some domains we consider inversion through the unit circle is not a symmetry, but the product of this inversion with a reflection is a symmetry. The discussion that follows also applies to this case.} We can think of a reflection as the limit of an inversion where $r_I \to \infty$, and so we see that there are no additional terms generated by this reduction (i.e. we can simply integrate over half the domain and multiply by a factor of $2$). Therefore reducing the domain by a product of $s$ reflections and an inversion through the unit circle yields
\ban{
\int_D d^2 w \left(\nabla \phi\right)^2 & = 2^{s+1}\left[\int_{\tilde D}d^2 w  \left(\nabla \phi\right)^2 + \int_{\tilde D}d^2 w  \left(\frac{2}{r}\partial_{r}\phi + \frac 2{r^2} \right)\right]\notag\\
& = 2^{s+1} \int_{\tilde D}d^2 w  \left(\nabla \phi\right)^2 + 2^{s+2}  \int_{\partial \tilde D} \phi \, d\theta + 2^{s+2} \int_{\partial \tilde D} \log r\,  d\theta \, .
}
We can additionally integrate by parts to get 
\ban{
\int_D d^2 w \left(\nabla \phi\right)^2 &  = -2^{s+1} \int_{\tilde D}d^2 w  \phi \nabla^2 \phi +2^{s+1} \int _{\partial \tilde D} \phi \nabla_n \phi  + 2^{s+2} \int_{\partial \tilde D} \phi \, d\theta + 2^{s+2} \int_{\partial \tilde D} \log r\,  d\theta \, ,\label{inversionCont}
}
which we can further simplify using the equations of motion $\nabla^2 \phi = e^{2\phi}$. 

Note that with our assumptions the boundary of $\tilde D$ consists of lines through the origin, the unit circle $U$, and some portion of the circles $C_i, C_i'$, and so we write $\partial \tilde D = \{\partial D_i\}$. It is thus convenient to write the integrals over $\partial \tilde D$ above as integrals over these lines and circles. All the integrals over the lines through the origin vanish due to either $d\theta$ or $\vec \nabla{}_n \phi$ vanishing, and so we are left with the integral over circles. Denoting $\mathscr I [\partial D_i]$ the contribution to the action from boundary segment $\partial D_i$, we can write the action as
\ban{
- \frac{24\pi}{c} I = - 4 \pi (g-1)(1-\log 4) - 2 \int_{\tilde D}\phi\,  e^{2\phi} \, d^2 w + \sum_i \mathscr I \left[\partial D_i\right]\, ,
}
where $\mathscr I[\partial D_i]$ includes possible contributions coming from eq. \eqref{TZaction}. We now compute this contribution for each type of boundary segment. The following subsection is rather technical, and should be thought of as a compendium of useful formulas. The reader more interested in the overall narrative should skip to the example in \S\ref{sec:example}. 

\subsection{Boundary circle contributions}

For simplicity, in this section we only compute the contribution reduced over inversion through the unit circle (or the product of this inversion and a reflection). Reducing over more reflections is straightforward and simply multiplies certain terms by factors of $2$. Throughout this section, we leave the sign inherited through the orientation of $\partial D$ implicit, i.e. we have
\ban{
\int_{C_i} d\theta^{(i)}_0 = \pm\,  2 \pi 
}
where we choose the positive or negative sign when $D$ lies inside or outside $C_i$ respectively. 

As previously mentioned, when $\partial D_i$ is a line the contribution $\mathscr I [\partial D_i]$ vanishes. For the unit circle $U$, we only have the contribution from eq. \eqref{inversionCont}: 
\ban{
\mathscr I [ U] = 2 \int_{U}  \phi \nabla_n \phi \, d\theta  + 4 \int_{U} \phi \, d\theta \, ,
}
as the $\log r$ term vanishes on $U$. We can use eq. \eqref{INVbc} to rewrite the normal derivative and we have 
\ban{
\mathscr I [U] = 2 \int_{U} \phi \, d\theta \, .
}
The rest of the boundary segments are made up of parts or all of $C_i$ and $C_i'$. There are multiple cases depending on the positions of these circles, and we go through all of them in detail. Note that we only consider cases in which the domain can be reduced by at least inversion through the unit circle, and additionally in which $C_i$ and $C_i'$ are the fixed point set of a symmetry of the domain.

In the simplest case, only one of $C_i$ or $C_i'$ is included in $\partial \tilde D$ and this circle does not intersect $U$. Without loss of generality, we can choose $C_i$ to be included in $\partial \tilde D$, so we have contributions to $\mathscr I [C_i]$ from eq. \eqref{inversionCont} and from the final summation in eq. \eqref{TZaction}. Using the boundary conditions eq. \eqref{FEMbc} we have
\ban{
\mathscr I[ C_i] = -2 \int_{ C_i} \phi\,  d\theta_0^{(i)} + 4 \int_{ C_i} \phi \, d \theta +4 \int_{ C_i} \log r d \theta + 4 \int_{ C_i} d\theta_\infty^{(i)} - 4 \pi \log |R_i ^2 - \Delta_i ^2 | \, ,
}
where $\theta_0^{(i)}$ is the angular coordinate measured from the center of $C_i$. Additionally, one can show 
\ban{
 \int_{ C_i} \log r d \theta  = \pi \log \left( 1- R_i^2/X_i^2\right)\, ,
}
for $X_i > R_i$ where $X_i$ is the Euclidean distance of the center of $C_i$ from the origin. Putting everything together we have
\ban{
\mathscr I[ C_i] = 2 \int_{ C_i} \phi\,  \left( 2 d \theta +2 d\theta_\infty^{(i)}- d\theta_0^{(i)}\right)  +4 \pi \log \frac{1- R_i^2/X_i^2}{|R_i ^2 - \Delta_i ^2 |} \, .
}
Further, numerically it is only convenient to integrate over $d\theta_0^{(i)}$, and so we can introduce Jacobian factors to transform $d\theta$ and $d\theta_\infty^{(i)}$. In general integrating on $C_i$ over an angle $\xi$ measured from a point along the axis connecting the origin and $X_i$ introduces the factor\footnote{Note using the signed Jacobian factor is more convenient numerically as a built-in way to keep track of possible orientation reversal.}
\ban{
\frac{d\xi}{d\theta_0^{(i)}} = \frac{R_i (R_i-d \cos \theta_0^{(i)})}{d^2-2 \, d \, R_i \cos \theta_0^{(i)}+R_i^2}\, , \label{Jangle}
}
where $d$ is the signed distance between $X_i$ and the point. For example, applying this formula to $\theta$ we have $d= - X_i$ and
\ban{
\frac{d\theta}{d\theta_0^{(i)}} = \frac{R_i (R_i+X_i \cos \theta_0^{(i)})}{X_i^2+2 \, X_i \, R_i \cos \theta_0^{(i)}+R_i^2}\, .
}

In the second case, we assume both $C_i$ and $C_{i'}$ are fully contained in $\partial \tilde D$. By the symmetry assumptions, there must be a conjugate pair $C_{\bar i}$, $C_{\bar i'}$ related by inversion through the unit circle. Therefore we must account for the contribution from this pair as well. Following similar arguments and using the transformation of $\phi$ under the inversion, we have
\ban{
\mathscr I [C_i ] + \mathscr I [C_i']  = &    2 \int_{C_i} \phi \,(2d\theta+2d\theta_\infty^{(i)}-d\theta_0^{(i)}) +4 \int_{C_i}(\phi +2 \log|w|) \frac{d\theta_\infty^{(\bar i)}}{d \theta_0^{(\bar i)}}\frac{d\theta_0^{(\bar i)}}{d \theta_0^{(i)}}  d\theta_0^{(i)}  \notag  \\
&+2 \int_{C_{i'}} \phi \,(2d\theta-d\theta_0^{(i')})+ 4\pi \log \frac{ (1-  {R_i^2}/{X_i^2})(1-  {R_{i'}^2}/{X_{i'}^2})}{\left |R_i^2 - \Delta_i^2 \right|\left |R_{\bar i}^2 - \Delta_{\bar i}^2 \right| } \, . 
}
We can similarly introduce Jacobian factors of the form eq. \eqref{Jangle} to numerically evaluate these integrals. The Jacobian for transforming the integral on $C_{(\bar i)}$ to one on $C_{(i)}$ can be worked out geometrically as
\ban{
\frac{d\theta_0^{(\bar i)}}{d \theta_0^{(i)}} = \frac{R_i^2 - X_i ^2}{{X_i^2+2 \, X_i \, R_i \cos \theta_0^{(i)}+R_i^2}} \, . \label{Jinv}
}

Finally, we have to consider the cases in which $C_i$ and $C_i'$ intersect the unit circle. First, we consider when $C_i$ is mapped to itself under inversion through the unit circle. In this case the analytic formulas were worked out in \cite{MRW} and we have
\ban{
\mathscr I[C_i] + \mathscr I[C_i'] = 2 \int _{\tilde C_i } \phi\, d\theta_0^{(i)}+2 \int _{\tilde C_i' } \phi\, d\theta_0^{(i')} - 8 \pi \log R_i + 8 \int_0 ^{2 \arctan R_i } \frac{x}{\sin x} dx\, ,
}
where $\tilde C_i$ refers to the part of $C_i$ that is part of $\partial \tilde D$ and similarly for $\tilde C_i'$. 

Additionally, we can consider the case when inversion through the unit circle is not a symmetry, but the product of this inversion and a reflection is a symmetry. In this case, the part of $C_i$ outside of $\tilde D$ gets mapped to the part of $C_{i'}$ inside $\tilde D$, and so we must include the appropriate Jacobian factor for inversion as in eq. \eqref{Jinv}, with an extra minus sign to account for the reversal of orientation. 
\ban{
\mathscr I [C_i ] + \mathscr I [C_i']  = &    2 \int_{\tilde C_i} \phi \,(2d\theta+2d\theta_\infty^{(i)}-d\theta_0^{(i)}) + 4 \int_{\tilde C_i} \log |w| \, d\theta  \notag  \\
&+2 \int_{\tilde C_{i'}} \phi \,(2d\theta-d\theta_0^{(i')}) +  4 \int_{\tilde C_{i'}} \log |w| \, d\theta \notag \\
& +4 \int_{\tilde C_{i'}}(\phi + 2 \log|w|)\frac{d\theta_\infty^{(i)}}{d \theta_0^{(i)}} \frac{d\theta_0^{(i)}}{d \theta_0^{(i')}}  d\theta_0^{(i')}- 4\pi \log {\left |R_i^2 - \Delta_i^2 \right|} \, . 
}

All of the above formulas are included in the attached \textit{Mathematica} package, providing a convenient set of tools to study these phases. 
%
%

\section{A \textit{Mathematica} package}
\label{sec:packages}

In this section, we document the usage of the attached \textit{Mathematica} package for computing the action and moduli of a handlebody phase. There are two packages included; \textit{FEMfine.m} implements general finite element methods for numerically solving differential equations, and \textit{handlebodies.m} provides a framework for solving for the handlebody geometry. 

To load the packages, make sure both files are included in the same directory as the working notebook and execute
\begin{verbatim}
SetDirectory[NotebookDirectory[]];
<<``handlebodies.m'';
\end{verbatim}
The \textit{FEMfine} package is automatically loaded as part of \textit{handlebodies}. 

\begin{figure}[h!]
\centering
\includegraphics[width=0.75\textwidth]{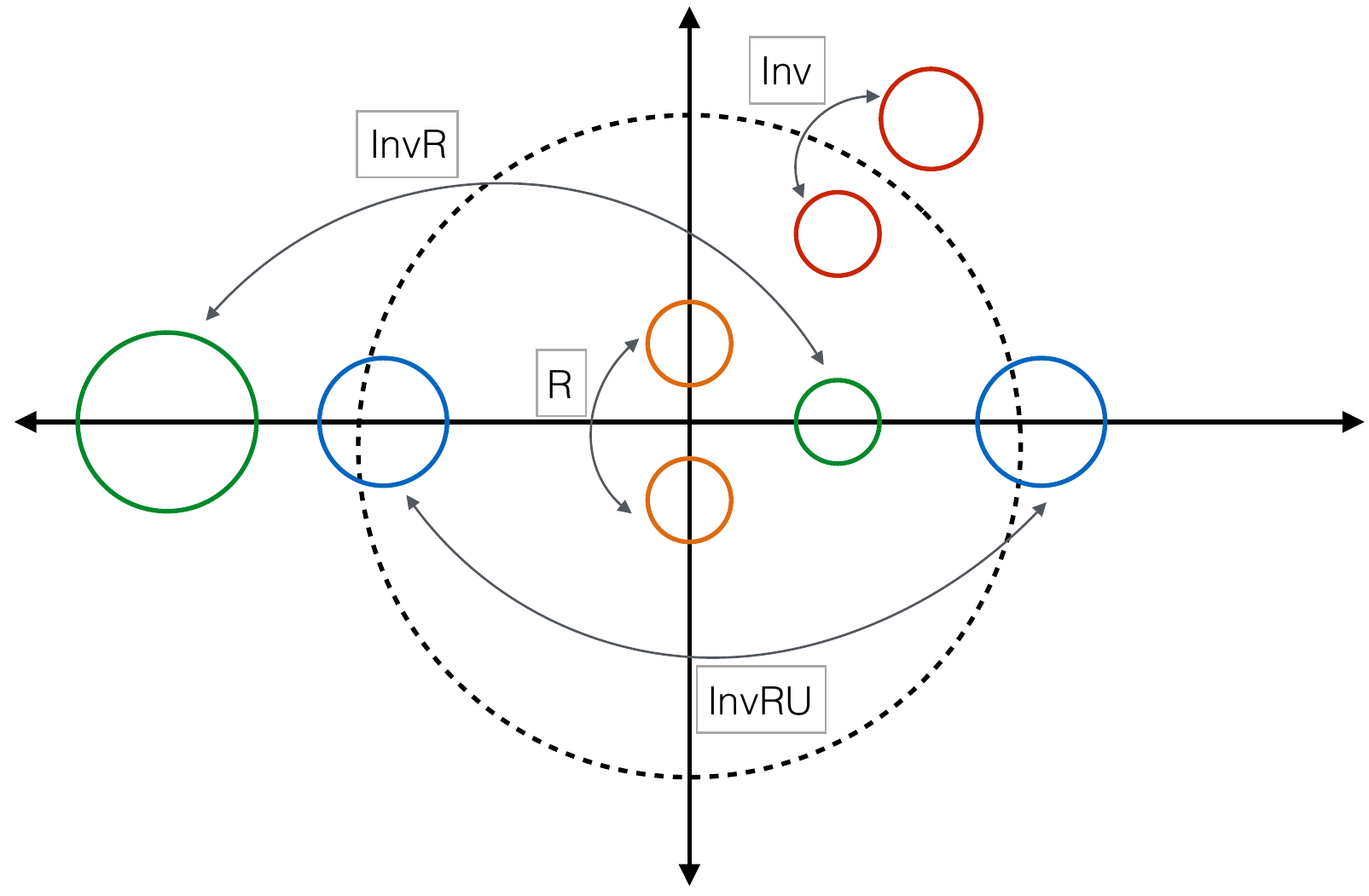}
\caption{Illustration of the allowed circle types according to the symmetries that exchange $C_i$ and $C_i'$ as follows. ``R": a reflection across $\hat x$ or $\hat y$. ``Inv:''  inversion through the unit circle. ``InvR:'' product of a reflection and inversion through the unit circle. ``InvRU:'' circles which are InvR and also intersect the unit circle. ``RU'' (not pictured): circles exchaged by reflection and also intersect the unit circle, and additionally must be fixed under inversion through unit circle.  \label{circletypes}
}
\end{figure}

To solve for a handlebody, one must first specify the circles $C_i$ and $C_i'$ in the domain and the symmetries. There are five allowed circle types, as documented in figure \ref{circletypes}, categorized according to the symmetry that $C_i$ and $C_i'$ are the fixed point set under.
Inversion in the unit circle must be a symmetry of the domain, and one can additionally speicfy reflection across the $x$ axis or $y$ axis as symmetries. This framework allows one to construct all of the handlebody phases considered in \cite{MRW,cones,toruspaper}, and additionally one can construct a large set of handlebodies for general application.

In the \textit{handlebodies} package, one can specify the handlebody via the following code:
\begin{verbatim}
InitializeHandlebody[]
AddCircle[{c1, r1, t1}]
AddCircle[{c2, r2, t2}]
...
AddSymmetry[``x'']
AddSymmetry[``y'']
\end{verbatim}
where $c=\{c_x,c_y\}$ is the center of each circle, $r$ is the radius, and $t$ is the type. The function \verb!IntializeHandlebody[]! resets the list of circles and symmetries, and sets the  mesh generation parameters in \textit{Mathematica}'s \verb!ToElementMesh! function as ``MaxCellMeasure''$\to 0.005$ and ``AccuracyGoal''$\to 4$. To increase the quality of the mesh, one can change the values of these parameters by resetting the variables \verb!mcm! and \verb!ag! to the desired values after the handlebody is initialized. 

Once the handlebody is specified, the executing the command \verb!SolveHandlebody[name]! computes a set of quantities and stores them as \verb!name[``Attribute'']!. If no variable \verb!name! is specified the attributes are stored as \verb!Handlebody[``Attribute'']!. The full list of quantities computed can be seen in the package documentation, and a few relevant ones are listed below.
\begin{itemize}
\item \verb!name[``genus'']!: Genus of boundary Riemann surface
\item \verb!name[``mesh'']!: Finite element mesh used to discretize domain $D$
\item \verb!name[``CError'']!: Estimation of numerical error due to discretization by mesh
\item \verb!name[``AError'']!: Estimation of numerical error from computation of area compared to the area determined from the genus by the Gauss-Bonnet theorem
\item \verb!name[``BoundaryLengths'']!: List of the length of each boundary segment $\partial D_i$ compute using the solution for the metric in the order \{circle segments, x segments, y segments\} with the order for the circle segments given by the order they were added, with the unit circle first.
\item \verb!name[``Action'']!: The Einstein-Hilbert action for the handlebody
\end{itemize}
One can read off various moduli of the Riemann surface in the list of boundary segment lengths, and additionally one can use this list to construct the analogous region in $\mathbb H^2$ to compute the rest of the moduli. 

Additionally, one must match moduli between different phases to determine the dominant phase for given boundary conditions. The \verb!NM! function and \verb!GradSearch! function are included as part of \textit{handlebodies} as convenient ways to match moduli using Newton's method and a gradient search method respectively. The documentation for these functions can also be read off from the package. 

%
%

\section{An example}

\label{sec:example}

\begin{figure}[h!]
\centering
\includegraphics[width=0.65\textwidth]{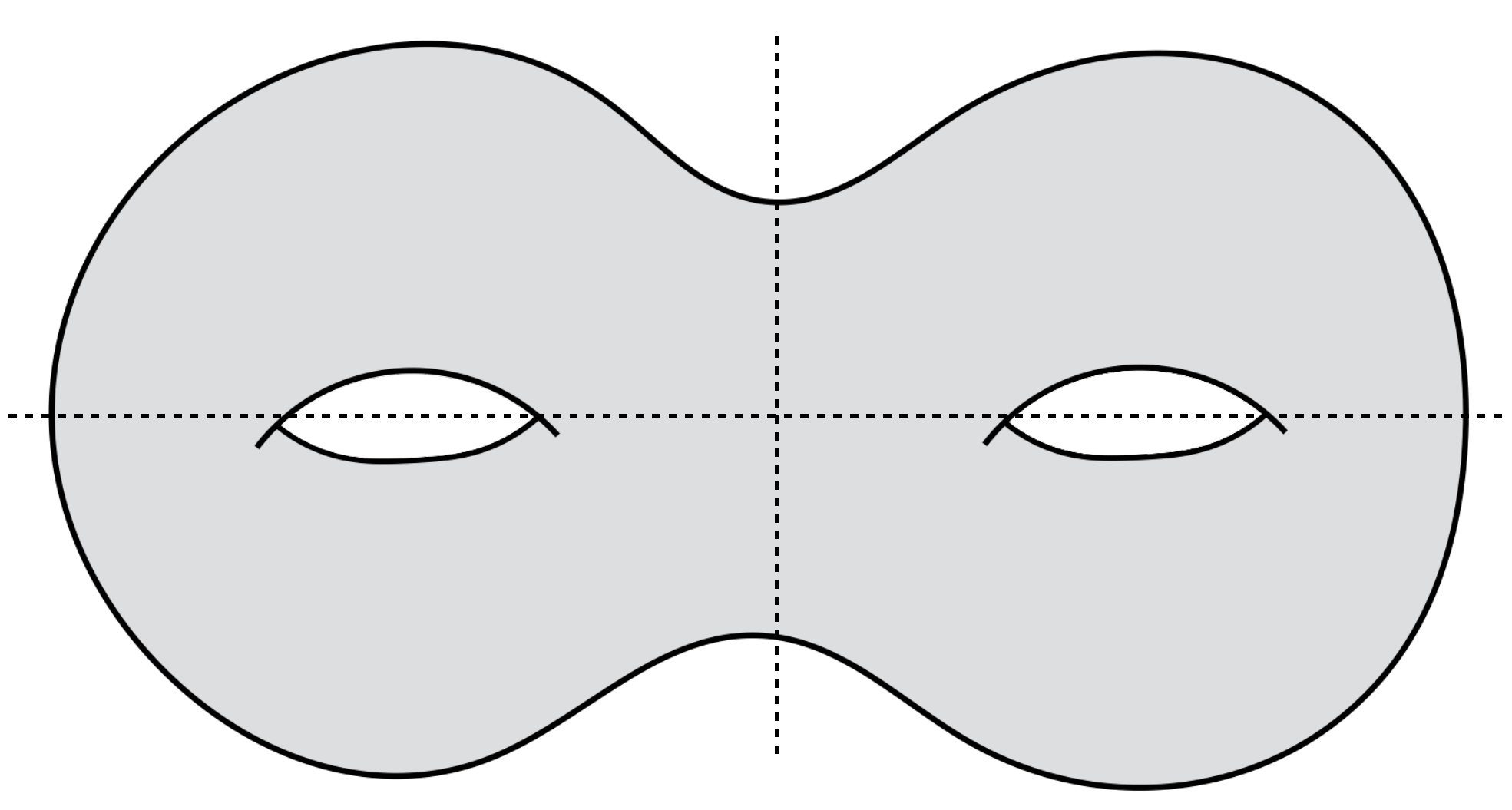}
\caption{Boundary Riemann surface with three $\mathbb Z_2$ symmetries given by reflection in each dashed line and the plane of the page.  \label{X2}
}
\end{figure}

As an example, we can use the \textit{handlebodies} package and the methods outlined in this section to study the toroidal geon phase originally studied in \cite{MRW}. First, we choose boundary conditions given by a genus $2$ Riemann surface drawn in figure \ref{X2} with three $\mathbb Z_2$ symmetries. 
This Riemann surface has a two dimensional moduli space leftover after imposing these symmetries. 

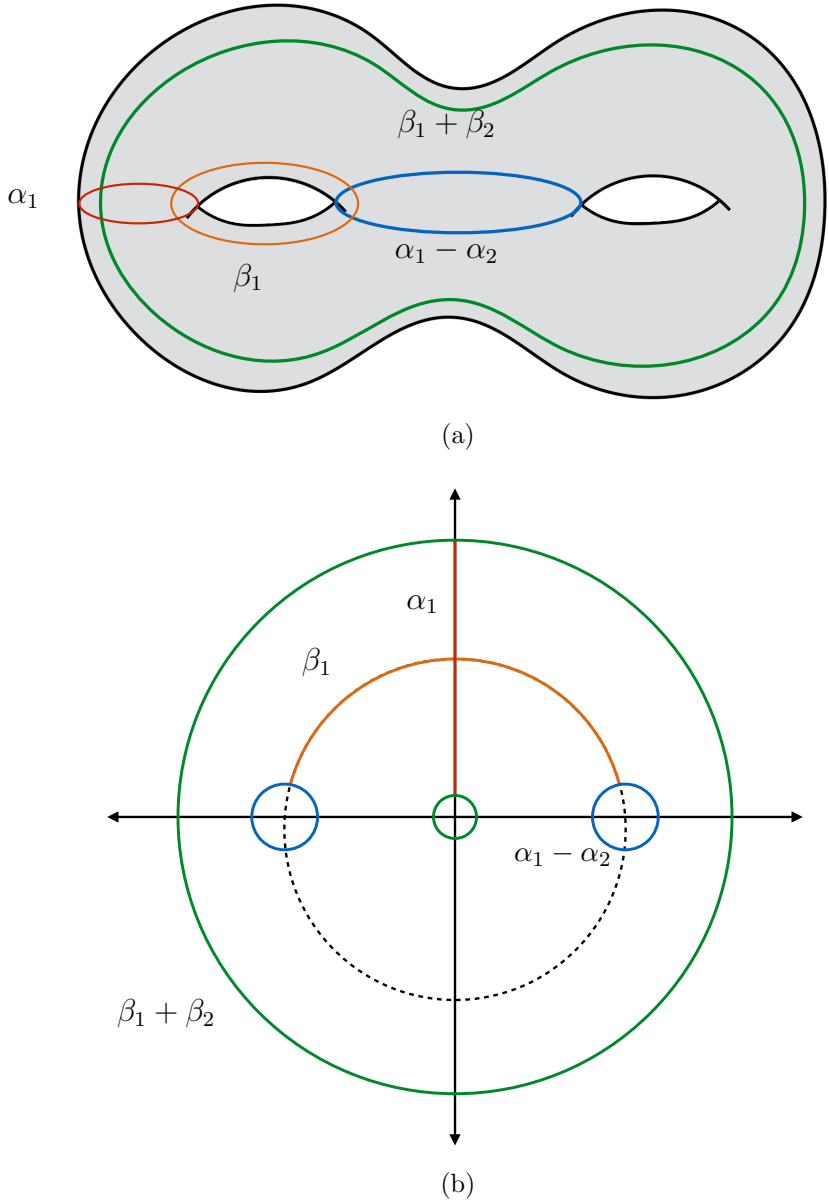
\begin{figure}[h!]
\centering
\subfloat[]{\includegraphics[width=0.65\textwidth]{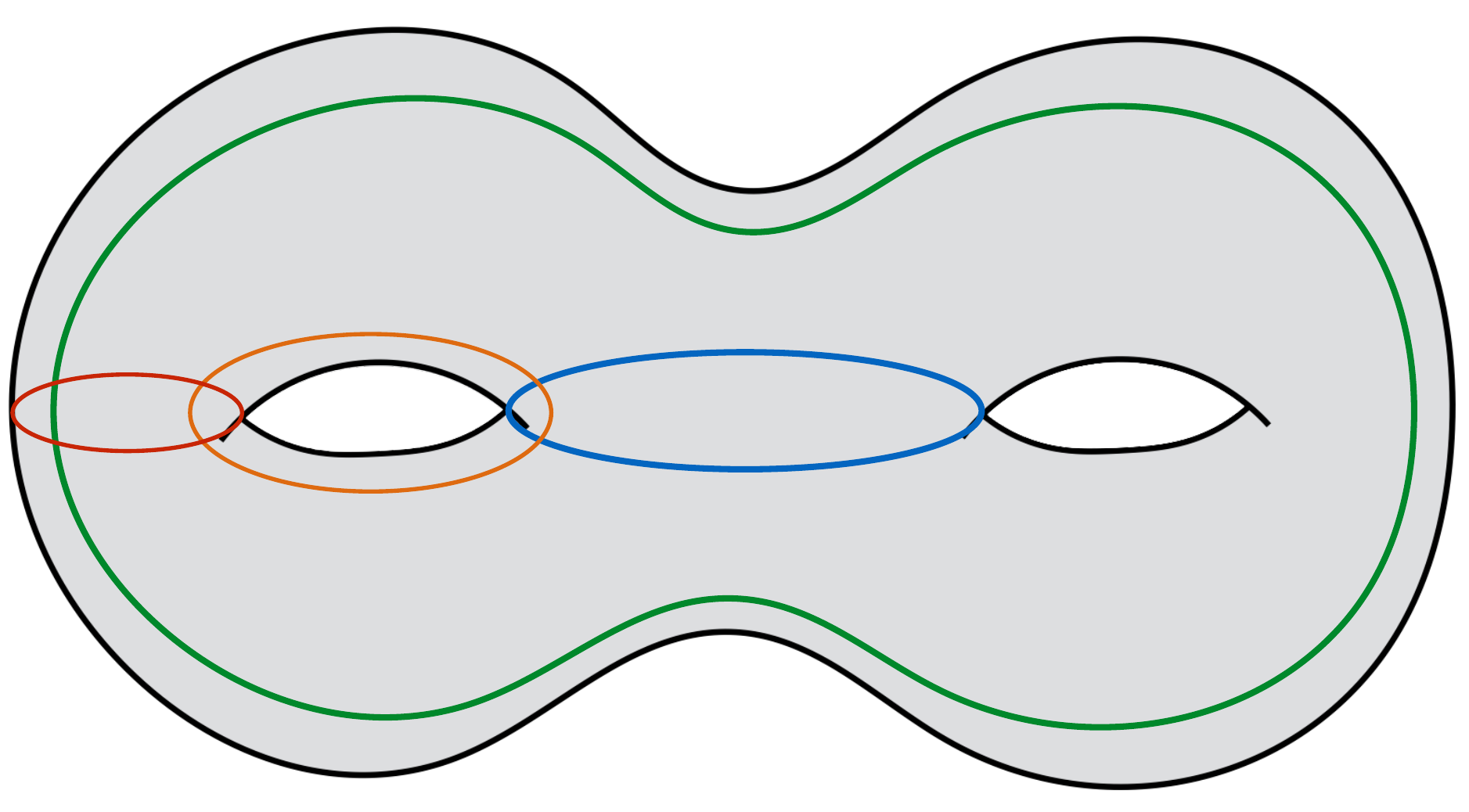}  \label{X2cycles}
\put(-310,80){\makebox(0,0){$\alpha_1$}}
\put(-225,50){\makebox(0,0){$\beta_1$}}
\put(-150,60){\makebox(0,0){$\alpha_1-\alpha_2$}}
\put(-150,108){\makebox(0,0){$\beta_1+\beta_2$}}
}\\
\subfloat[]{\includegraphics[width=0.6\textwidth]{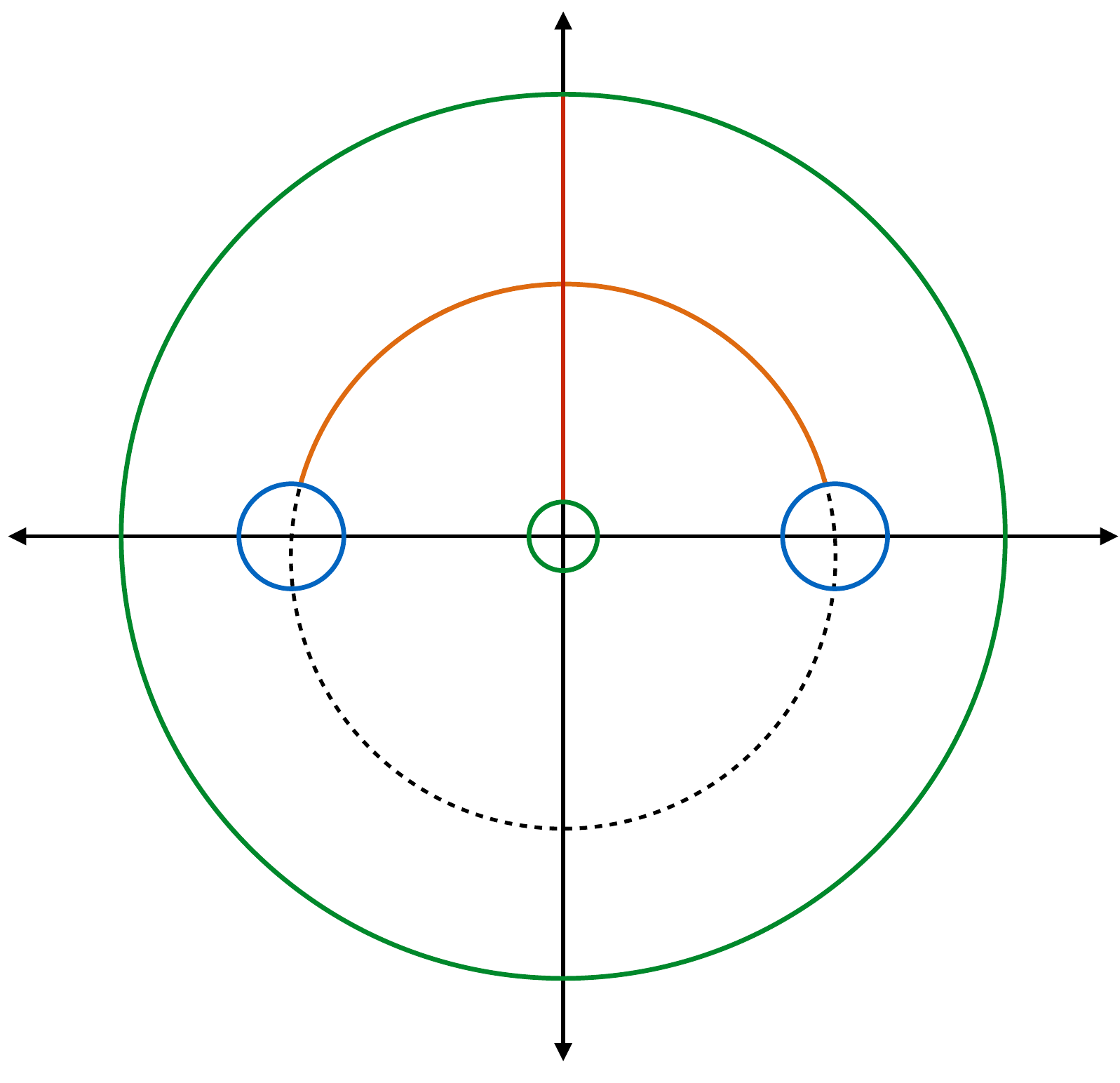} \label{X2domain}
\put(-148,210){\makebox(0,0){$\alpha_1$}}
\put(-188,188){\makebox(0,0){$\beta_1$}}
\put(-95,115){\makebox(0,0){\small$\alpha_1-\alpha_2$}}
\put(-245,55){\makebox(0,0){$\beta_1+\beta_2$}}
}
\caption{(a) Cycles labeled on the boundary Riemann surface. (b) The Schottky uniformization of this Riemann surface used to compute the toroidal geon phase. 
}
\end{figure}

In order to specify a handlebody phase, we choose two independent cycles to make contractible in the bulk. There are three distinct choices that respect the $\mathbb Z_2$ symmetries of the boundary. Letting the $\alpha$ cycles go around the handles (red in fig \ref{X2cycles}) and the $\beta$ cycles go around the holes (orange in fig \ref{X2cycles}), the phases are defined by choosing $\{\alpha_1, \alpha_2\}$ contractible, choosing $\{\beta_1, \beta_2\}$, or choosing $\{\alpha_1 - \alpha_2, \beta_1+\beta_2\}$. Each of these choices results in a different handlebody phase. 

We choose to study the phase in which $\{\alpha_1 - \alpha_2, \beta_1+\beta_2\}$ are contractible. These cycles are drawn in blue and green on figure \ref{X2cycles} respectively. To study this phase we must first cut the Riemann surface apart along theses cycles and project it into the complex plane. First, cutting the Riemann surface along $\alpha_1-\alpha_2$ yields a square torus with two punctures related by a reflection symmetry. Next, we can cut this torus along its $\beta$ cycle (i.e. $\beta_1+\beta_2$) to yield the Riemann sphere with four punctures, where the punctures are identified by orthogonal reflection symmetries. Projecting this sphere into the plane gives the Schottky domain drawn in figure \ref{X2domain}. Reflection about the $\hat x$ and $\hat y$ axis identify each pair of $C_i$, $C_i'$, and inversion in the unit circle leaves the domain unchanged. Additionally, we can identify the cycles $\alpha_1$ and $\beta_1$ in this domain as the fixed point sets under the relevant symmetries. 

We can characterize the bulk geometry of this handlebody by considering the geometry of a particular time slice. Consider the slice given by the fixed point set of reflection across the vertical line in fig \ref{X2}. This symmetry fixes the $\hat x$ axis of the Schottky domain in fig \ref{X2domain}, and the topology of this slice is determined by eq. \eqref{sliceG}. The slice has $2$ pairs of circles, and the boundary consists of a single segment, giving a topology of $g_\text{slice}=1$. Therefore, in this phase this bulk time slice has geometry given by a single boundary wormhole with a genus one surface behind the horizon. As in \cite{toruspaper} we refer to this phase as the toroidal geon. 

Note that we could have chosen a different time slice to characterize the bulk geometry. A potential source of confusion is that doing so does not change the handlebody phase, but rather simply the bulk slice we are using to characterize it. If we had chosen either of the two remaining slices fixed by $\mathbb Z_2$ symmetries we would have resulted in a geometry with three boundaries-- with two of the boundaries connected by a wormhole, and the third a copy of the Poincar\'e disk. In each of these cases it is important to take the entire fixed point set of the reflection symmetry as the boundary. For example, if we considered the fixed point set of the reflection across the horizontal line in fig. \ref{X2} the boundary slice consists not only of the $\hat y$ axis but also of the cycle $\alpha_1-\alpha_2$. This statement is clear in fig. \ref{X2cycles} but more sublte in fig. \ref{X2domain}. 

\begin{figure}[h!]
\centering
\includegraphics[width=0.35\textwidth]{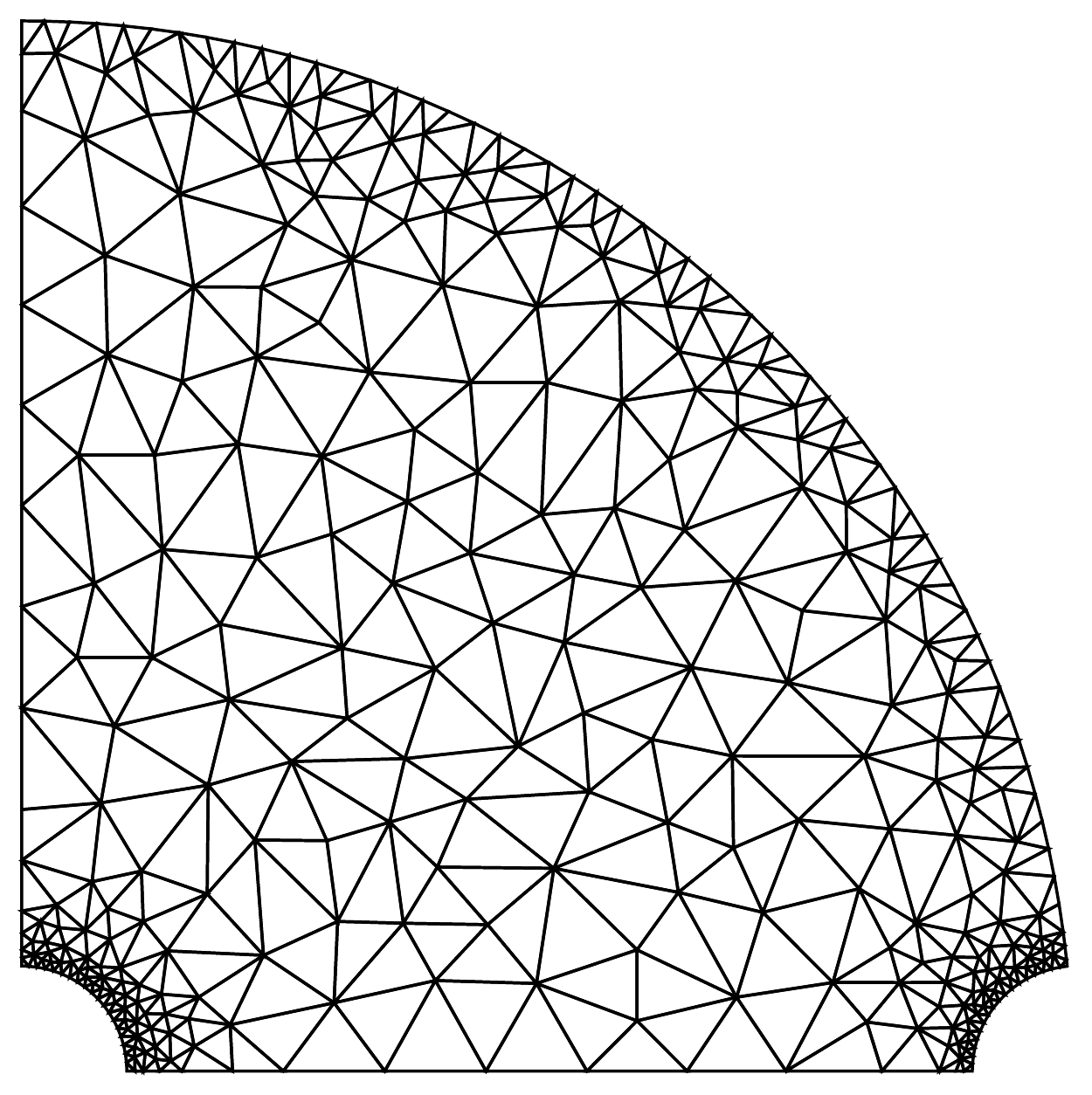}
\caption{A mesh used to compute the toroidal geon phase. \label{Xmesh}
}
\end{figure}

\begin{figure}[h!]
\centering
\includegraphics[width=0.6\textwidth]{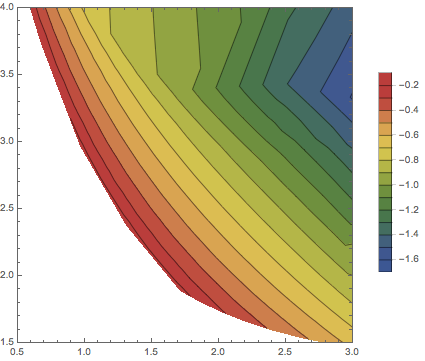}
\put(-20,0){\makebox(0,0){$|\alpha_1|$}}
\put(-285,225){\makebox(0,0){$|\beta_1|$}}
\caption{The action $I/c$ for the toroid geon as a function of moduli. We see the action decreases as $|\alpha_1|$ and $|\beta_1|$ increase. \label{Xaction}
}
\end{figure}

Having specified the phase, we can now compute its action and moduli using the \textit{handlebodies} package. We can construct a general such phase via the following code: 
\begin{verbatim}
InitializeHandlebody[]

AddCircle[{{0, 0}, r, ``Inv''}]
AddCircle[{{Sec[a], 0}, Tan[a], ``RU''}]

AddSymmetry[``x'']; AddSymmetry[``y''];

SolveHandlebody[geon]
\end{verbatim}
A sample mesh used for this phase is shown in figure \ref{Xmesh}. 
Evaluating this code for different values of $r$ and $a$ computes the action of this phase at various points in moduli space. To parameterize the moduli space, we can use $|\alpha_1|$ and $|\beta_1|$, which after reducing the Schottky domain by the three $\mathbb Z_2$ symmetries correspond to boundary segments $\partial D_i$. In figure \ref{Xaction} we show a contour plot of the action in this moduli space. 

\section*{Acknowledgments}
Thank you to Don Marolf, Henry Maxfield, and Benson Way for useful conversations. This work was supported in part by the U.S. National Science Foundation under grant number PHY15-04541 and also by the University of California.

\bibliographystyle{jhep}
\phantomsection
\renewcommand*{\bibname}{References}

\bibliography{references}

\end{document}